\documentclass[aps, pra, preprint, groupedaddress, amsfonts, longbibliography,
               amsmath, amssymb, showpacs, nofootinbib]{revtex4-1}

\usepackage{microtype}
\usepackage{graphicx}
\usepackage{epstopdf}
\usepackage[utf8]{inputenc}
\usepackage[T1]{fontenc}
\usepackage[usenames,dvipsnames]{xcolor}
\usepackage{hyperref}

\usepackage{color}

\begin{document}
\title{Proton-impact-induced electron emission from biologically relevant
molecules studied with a screened independent atom model}

\author{Hans J\"urgen L\"udde}
\email[]{luedde@itp.uni-frankfurt.de}
\affiliation{Frankfurt Institute for Advanced Studies (FIAS), D-60438 Frankfurt, Germany} 

\author{Marko Horbatsch}
\email[]{marko@yorku.ca}
\affiliation{Department of Physics and Astronomy, York University, Toronto, Ontario M3J 1P3, Canada}

\author{Tom Kirchner}  
\email[]{tomk@yorku.ca}
\affiliation{Department of Physics and Astronomy, York University, Toronto, Ontario M3J 1P3, Canada}
\date{\today}
\begin{abstract}
We use the recently introduced independent-atom-model pixel counting method to calculate proton-impact
net ionization cross sections for a large class of biologically relevant systems including
pyrimidines, purines, amino acids, and nucleotides from 10 keV to 10 MeV impact energy. 
Overall good agreement with experimental data, where available, is found.
A scaling prescription that involves coefficients derived from the independent atom model is shown to
represent the cross section results better than scalings based on the number of (bonding)
valence electrons of the target molecules. 
It is shown that the scaled net ionization cross sections  
of the proton-nucleotide collision systems can be represented in terms of a simple analytical formula
with four parameters
to within 3\% accuracy.
 
\end{abstract}

\maketitle
\section{Introduction}
\label{intro}
There has been growing interest in collisions involving organic compounds and biomolecules in recent years.
That interest and the ensuing research activity are largely driven by data needs in areas ranging from
astrochemistry to ion-beam cancer therapy. 
Ideally, the cross section data required, e.g., for
a detailed understanding of the radiation damage of biological tissue~\cite{Garcia12}, 
would be obtained from systematic measurements and first-principles calculations.
While progress has been made on both 
experimental (see~\cite{Moretto06, Tabet10a, Tabet10b, Iriki11a, Iriki11b, Itoh13, Wolff14, Rudek16} 
for proton-impact collisions) and 
theoretical~\cite{Montabonel16, Covington17, Teixeira18} fronts,
the complexity and multitude of molecules of interest
suggest that 
there is a role to be played by simplified
models which are easily applicable to a wide range of collision systems.

Classical arguments or quantum-mechanical first-order perturbation
theory represent natural starting points
for modeling electron removal in ion-molecule collisions.
Making use of these ingredients,
a number of
analytical cross section formulae have been proposed
and applied~(see, e.g.,~\cite{Rudd85a, Stolterfoht97, Montenegro13} and the discussion
in~\cite{Rudek16}).
Their advantage is simplicity---typically they only require the binding energies
of the target electrons as physical input---but their success depends on (additional) parameters which
are determined (semi-) empirically. As a consequence, these analytical models do not have sufficient
predictive power, and are thus of
limited value for problems for which measurements or more sophisticated calculations are 
not available.

Most attempts at constructing more sophisticated models are based on the idea that the
ionization or capture cross section of a complex molecule may be linearly combined from smaller
parts. In the {\it complete neglect of differental overlap} (CNDO) approach a
Mulliken-like population analysis is applied to an
electronic structure calculation of the target molecule.
The molecular cross section is then written as a linear combination of contributions from
all atomic orbitals involved, with
the Mulliken populations as weight factors~\cite{Champion10}.  
A related class of models is based on the additivity rule (AR) according to which the ionization
or capture cross section
for a specifc target system is obtained as a sum of cross sections of its building blocks. 
Normally, atomic building blocks are used and the description is referred to as 
independent-atom-model (IAM)-AR. 
In a recent work it was argued that one can also start
from small molecular constituents to assemble the ionization cross section of a 
larger molecule~\cite{Paredes15}. A caveat of this independent-molecule-model (IMM)-AR is that 
there are usually many different ways to decompose a
given molecule into molecular building blocks and depending on which ansatz is used the
resulting cross section may vary. 
It should also be noted that the IMM-AR work of~\cite{Paredes15} used experimental 
cross section data for the building blocks as input, while the atom-like contributions in the CNDO 
approach are usually obtained from perturbative collision 
calculations~\cite{Champion10, Galassi12, Champion12, Champion13}.

Motivated by the somewhat limited scope of these models
we recently introduced an IAM-based
description of ion-molecule collisions which uses first-principles-based
atomic cross section calculations and
goes beyond the simple AR~\cite{hjl16}. 
The main assumption of our model is that the net ionization and capture cross
sections can be represented as 
{\it weighted} sums of atomic net cross sections for all the atoms that make up the molecule.
The weight factors account for the geometrical
overlap which occurs when one
projects the loci of the atomic centers of the molecule onto a
plane perpendicular to the projectile beam direction and 
pictures the atomic cross sections as circular disks in that plane. 
The ``visible'' effective cross sectional area is calculated
using a pixel counting method (PCM) and, accordingly, we refer to the
model as IAM-PCM. In the limit of small overlaps the IAM-PCM cross section approaches the result of
the IAM-AR. 
This is a desirable property since the IAM-AR is known to give fairly accurate
results in the high-impact-energy regime in which capture and ionization cross sections are small
and geometrical overlap is negligible.
Toward lower energies, capture and ionization become stronger and the IAM-AR tends to overestimate 
experimental data~\cite{Paredes15, hjl16}. 
In applications to proton collisions from
medium-sized molecules such as H$_2$O and from larger water, neon, and
carbon clusters 
we have found that geometrical overlap does 
occur at these lower energies and leads
to significant cross-section reductions~\cite{hjl16,hjl18}. 
Compared to the IAM-AR the agreement with experimental data,
where available, is improved.

Given these promising results and the relative simplicity of the model, the IAM-PCM seems
ideally suited to study collisions from complex
biomolecules for which neither ab-initio calculations nor 
measured cross section data
are available.
This is the main objective of the present work.
In particular, we examine scaling relations obtained from IAM-PCM calculations
for proton-impact net ionization cross sections
of different groups of systems such as amino acids and nucleotides, 
and suggest a parametrization of our results in terms of a simple
analytical formula.

The layout of the paper is as follows. We begin in section~\ref{sec:pcm} with a short
summary of the IAM-PCM. For a more detailed description the reader is referred to~\cite{hjl18}. 
In section~\ref{sec:validate} we seek to further validate the model, beyond the results presented
in~\cite{hjl16, hjl18}, by
comparing calculated net ionization cross sections with
available measurements and previous theoretical and semi-empirical model data for 10~keV to 10~MeV proton
collisions from 
pyrimidine (C$_4$H$_4$N$_2$), purine (C$_5$H$_4$N$_4$), 
tetrahydrofuran (THF -- C$_4$H$_8$O) and trimethyl phosphate (TMP -- (CH$_3$)$_3$PO$_4$).
Different scaling prescriptions for the net ionization cross section are examined and applied to
a large class of systems including amino acids and nucleotides in section~\ref{sec:scaling} and
a simple parametrization of these results is suggested in section~\ref{sec:para}.
The paper ends with concluding remarks in section~\ref{sec:conclusions}.

\section{Theoretical model}
\label{sec:pcm}
The ingredients of the IAM-PCM are atomic cross sections and molecular ground-state geometries. 
The latter are taken from the literature using a Cartesian coordinate representation~\cite{Young01}, 
which is commonly referred to as xyz-file format.
More specifically, for the biomolecules studied in this work we use
data provided through~\cite{molview}.
The proton-atom net ionization cross sections are calculated in a density functional
theory (DFT)
framework on the level of the independent
electron model. 
We use a well-tested no-response model in which the Kohn-Sham potential is
approximated by an accurate 
exchange-only ground-state potential~\cite{tom98} and time-dependent screening and exchange effects 
are neglected. 
The nonperturbative two-center basis generator method 
(TC-BGM) 
is used for orbital propagation~\cite{tcbgm}. 
It yields transition probabilities and cross sections for target excitation and
electron transfer to the projectile (capture) and the continuum (ionization). 
In the present work we only look at the ionization channel, since it usually exhibits simpler
scaling properties and there are no
capture data available for comparison for most of the molecules studied here.

The net ionization cross sections $\sigma_j^{\rm net}$ for the $j=1,\ldots , N$ atoms that make up the
molecule under study are combined according to
\begin{equation}
 \sigma_{\rm mol}^{\rm net}(E,\alpha,\beta,\gamma) = 
  \sum_{j=1}^N s_j(E,\alpha,\beta,\gamma)  \sigma_j^{\rm net} (E) 
\label{eq:scar-pcm}
\end{equation}
to yield the molecular cross section at projectile energy $E$. The weight factors $0\le s_j \le 1$
account for the overlap of the atomic cross sections and
depend on the 
relative orientation of the molecule with respect to the ion beam direction. This dependence is
captured by the Euler angles $\alpha , \beta , \gamma$. 

To calculate the weights we picture the atomic cross sections as circular disks with radii
$r_j(E) = [\sigma_j^{\rm net}(E)/\pi]^{1/2}$ in a plane perpendicular to the ion beam axis.
The combined area of overlapping circles is broken up into pixels and
calculated by counting those pixels that are visible to the impinging projectile.
Accordingly, the screening coefficients can be determined by
\begin{equation}
  s_j(E,\alpha,\beta,\gamma) = \frac{\sigma_j^{{\rm vis}} (E,\alpha,\beta,\gamma)}
                   { \sigma_j^{\rm net} (E)} ,
\label{eq:scar-coeff}
\end{equation}
where $\sigma_j^{{\rm vis}}$ is the visible part of the $j$th atomic cross section.

The IAM-PCM is similar in spirit to the so-called screening-corrected additivity rule (SCAR) developed
and used for electron-molecule collisions~\cite{Blanco2003}. SCAR cross sections, however, are
based on a heuristic recurrence relation for the determination of the
screening coefficients in an orientation-independent version of~(\ref{eq:scar-pcm}), whereas
the IAM-PCM procedure to calculate them for any given orientation is numerically exact.
In order to compare IAM-PCM calculations with 
experimental data for \textit{randomly} oriented molecules we repeat the pixel count for 
an ensemble of Euler angle triples and average the cross-section results appropriately.

Once the atomic cross sections have been calculated, the IAM-PCM
procedure is not at all resource intensive. It takes just a few minutes on a single-core
computer to carry out an orientation average at a given energy for a system consisting of
dozens of nuclei and hundreds of electrons.  
The reader is referred to~\cite{hjl18} for more details.

\section{Validation of the model}
\label{sec:validate}
The molecules pyrimidine, purine, 
THF, and TMP are structural analogues of 
DNA constituents and have been 
studied in recent collision experiments~\cite{Wolff14, Rudek16, Bug17}, since they
were deemed more amenable to gas-phase cross-section measurements 
than actual DNA building 
blocks\footnote{%
While it is difficult to prepare well-characterized gas targets
of large neutral DNA components, the technique of electrospray ionization offers a pathway to 
bringing {\it charged} complex biomolecules into the gas phase and use
them in collision experiments~\cite{Lalande18}.}.
More specifically, pyrimidine is a single carbon-nitrogen-ring molecule from which the nucleobases cytosine and
thymine (and the RNA nucleobase uracil) are derived, while the double-ring molecule purine is a precursor
of the nucleobases adenine and guanine. Both pyrimidine and purine are also used as generic names for 
wider classes of similar one-ring and two-ring molecules, respectively, which include the DNA and RNA nucleobases 
(see section~\ref{sec:scaling}).
THF serves as a model for the monosaccharide deoxyribose in DNA, 
and TMP represents the phosphate residue which together with the sugar molecule forms
the DNA backbone~\cite{Adams81, Alberts15}.

\begin{figure*}
\begin{center}
\resizebox{0.235\textwidth}{!}{%
\includegraphics{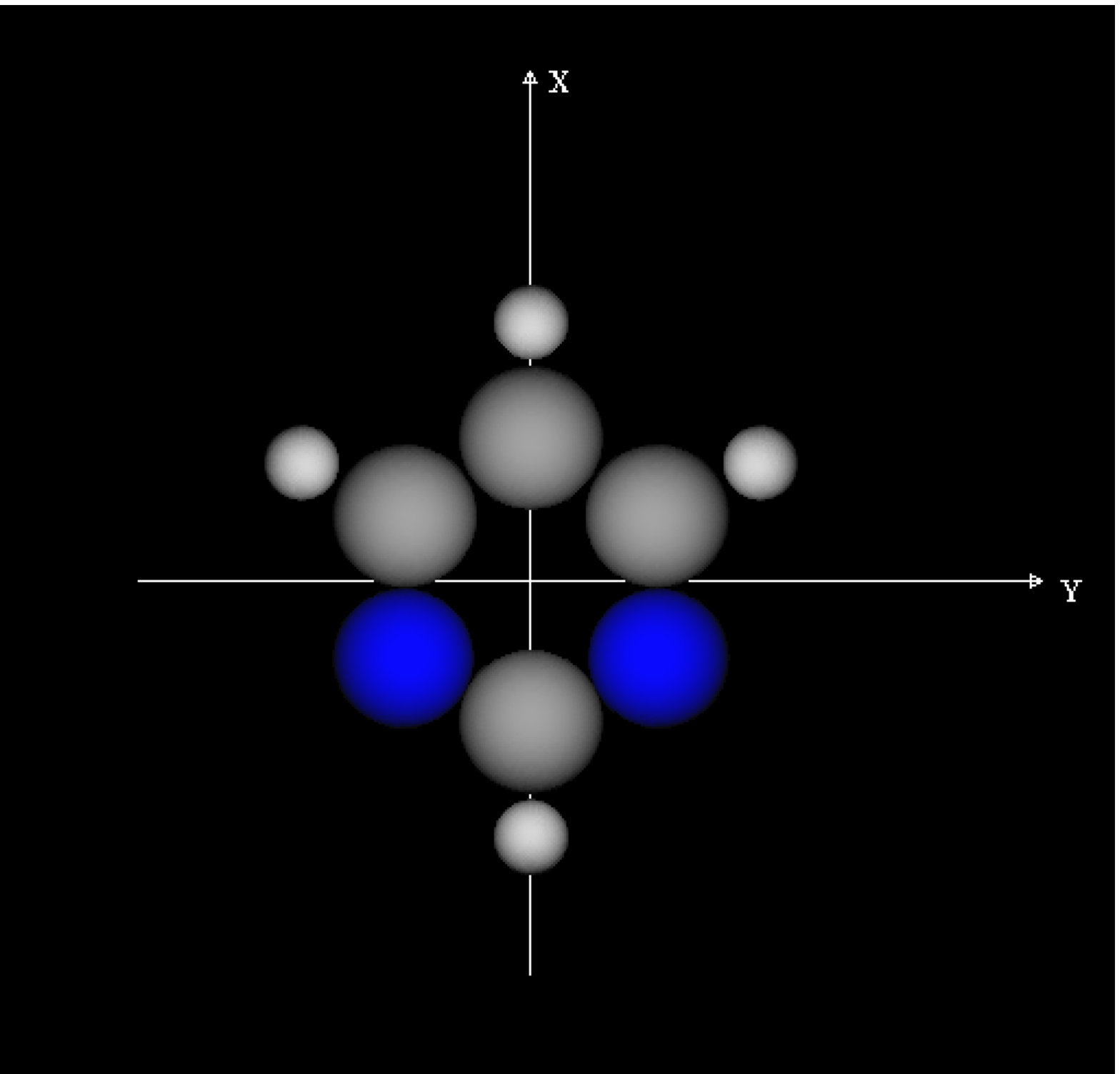}
}
\resizebox{0.229\textwidth}{!}{%
\includegraphics{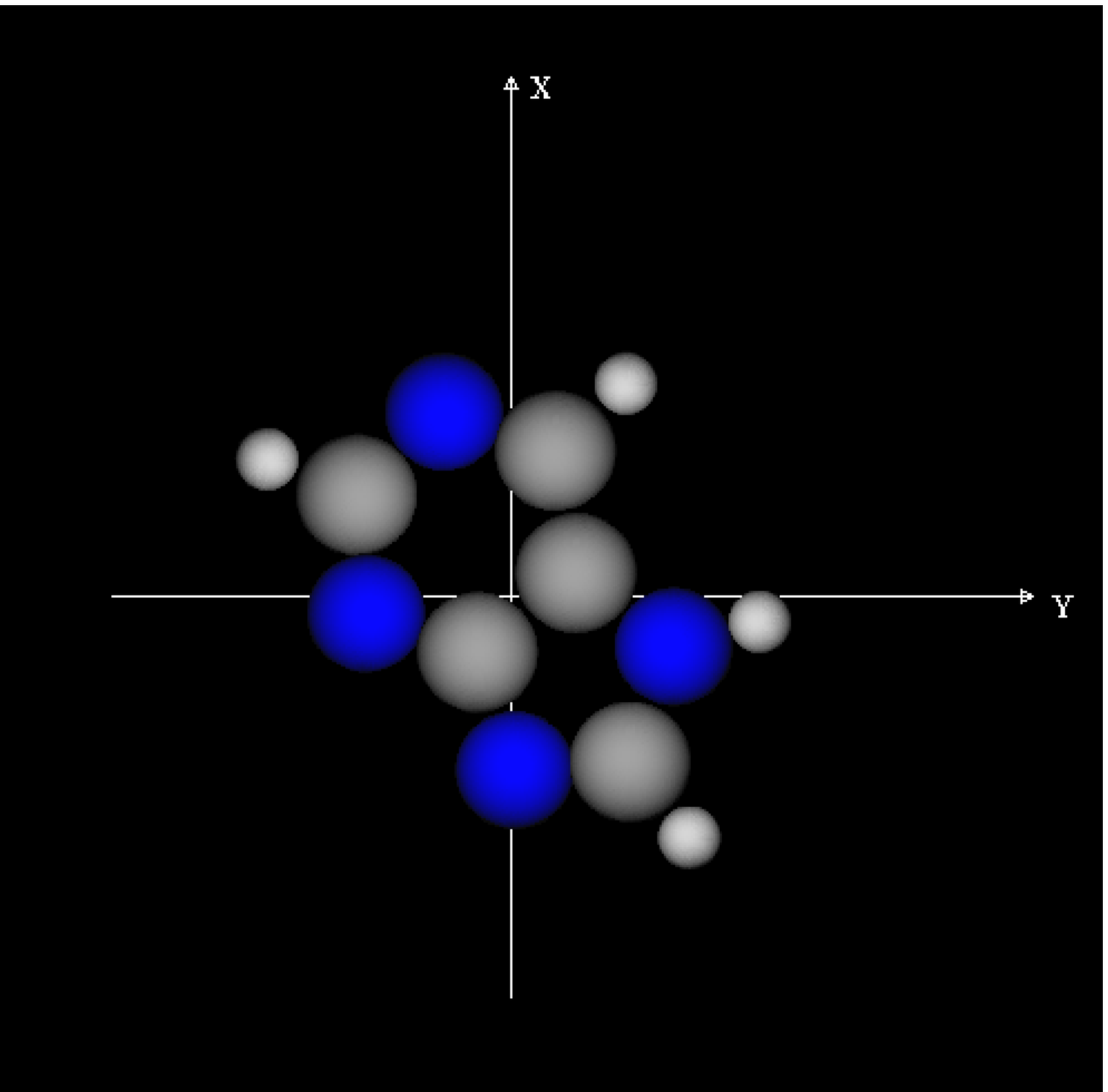}
}
\resizebox{0.242\textwidth}{!}{%
\includegraphics{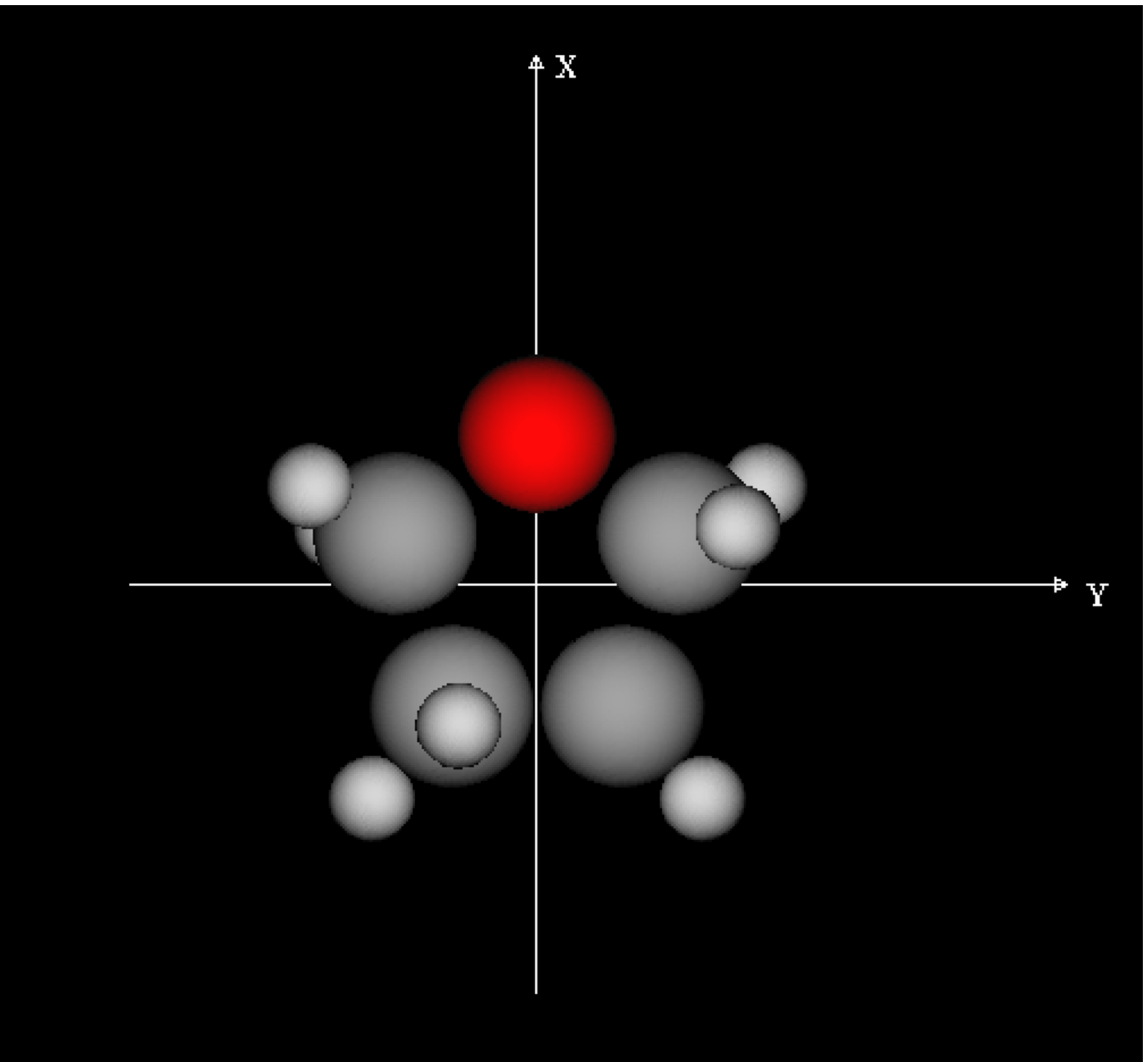}
}
\resizebox{0.242\textwidth}{!}{%
\includegraphics{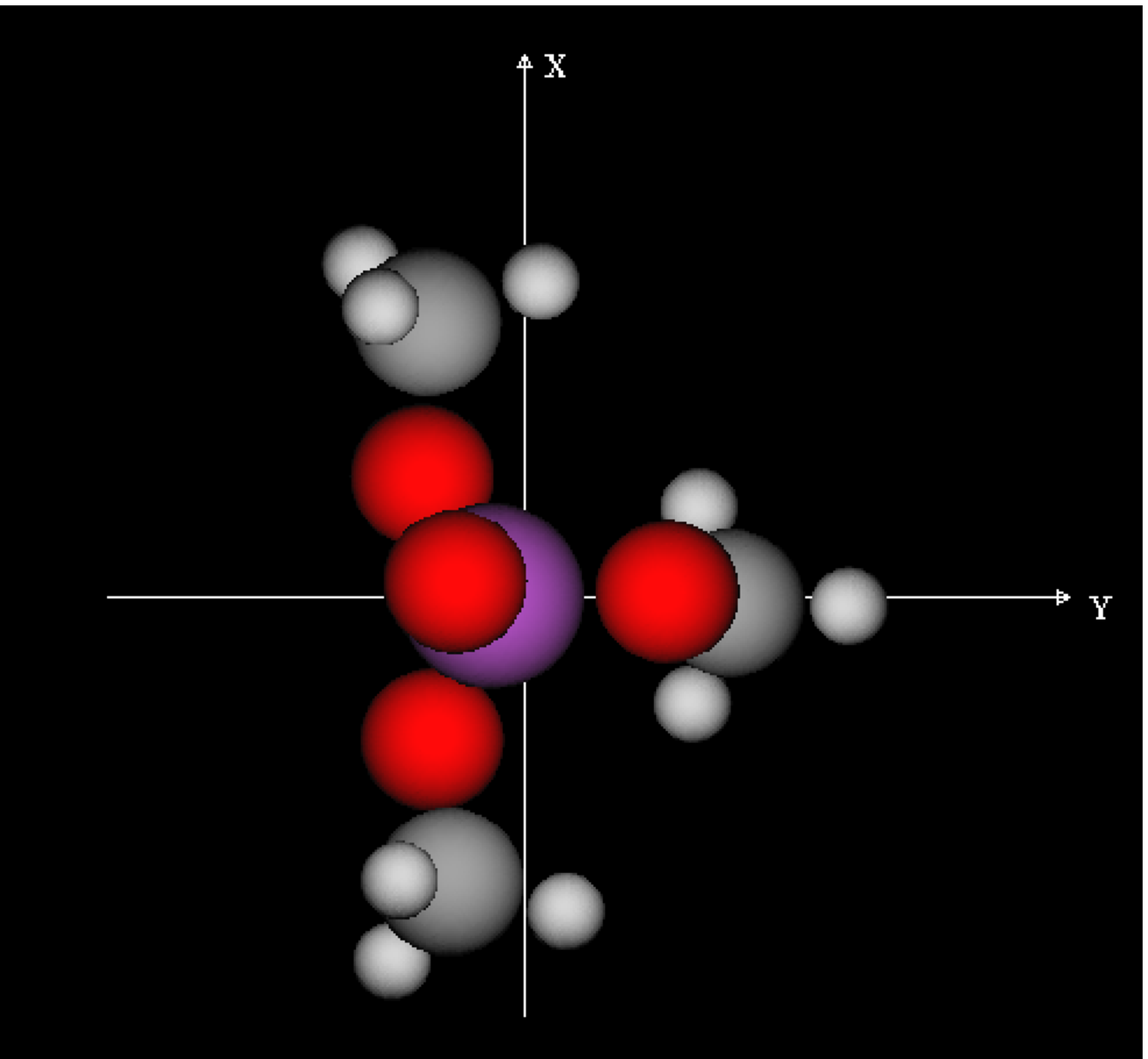}
}

\caption{%
Net ionization in $E=500$ keV proton collisions with (from left to right) pyrimidine (C$_4$H$_4$N$_2$),
purine (C$_5$H$_4$N$_4$), THF (C$_4$H$_8$O), TMP ((CH$_3$)$_3$PO$_4$) for particular orientations.
The radii of the circular disks are given by
$r_j = [\sigma_j^{\rm net}/\pi]^{1/2}$ using TC-BGM proton-atom net ionization
cross sections $\sigma_j^{\rm net}$.}
\label{fig:structures}
\end{center}
\end{figure*}

Figure~\ref{fig:structures} displays effective cross sectional areas for
proton collisions from these four
molecules for arbitrary orientations with respect to the projectile beam axis. The atomic
nuclei are placed at their equilibrium ground-state positions
and the plots are obtained from net ionization calculations at $E=500$ keV impact energy.
For pyrimidine and purine there is no overlap of the atomic cross
sections for the chosen geometries, while a modest overlap effect occurs for THF and
a somewhat larger one for TMP. 
One can imagine how the magnitude of the overlap effect varies as a function of 
orientation and collision energy and
that even for pyrimidine, the smallest of the four molecules, the orientation-averaged
IAM-PCM
cross section will be smaller than the 
zero-overlap limit corresponding to the IAM-AR. 

\begin{figure*}
\begin{center}
\resizebox{0.49\textwidth}{!}{%
\includegraphics{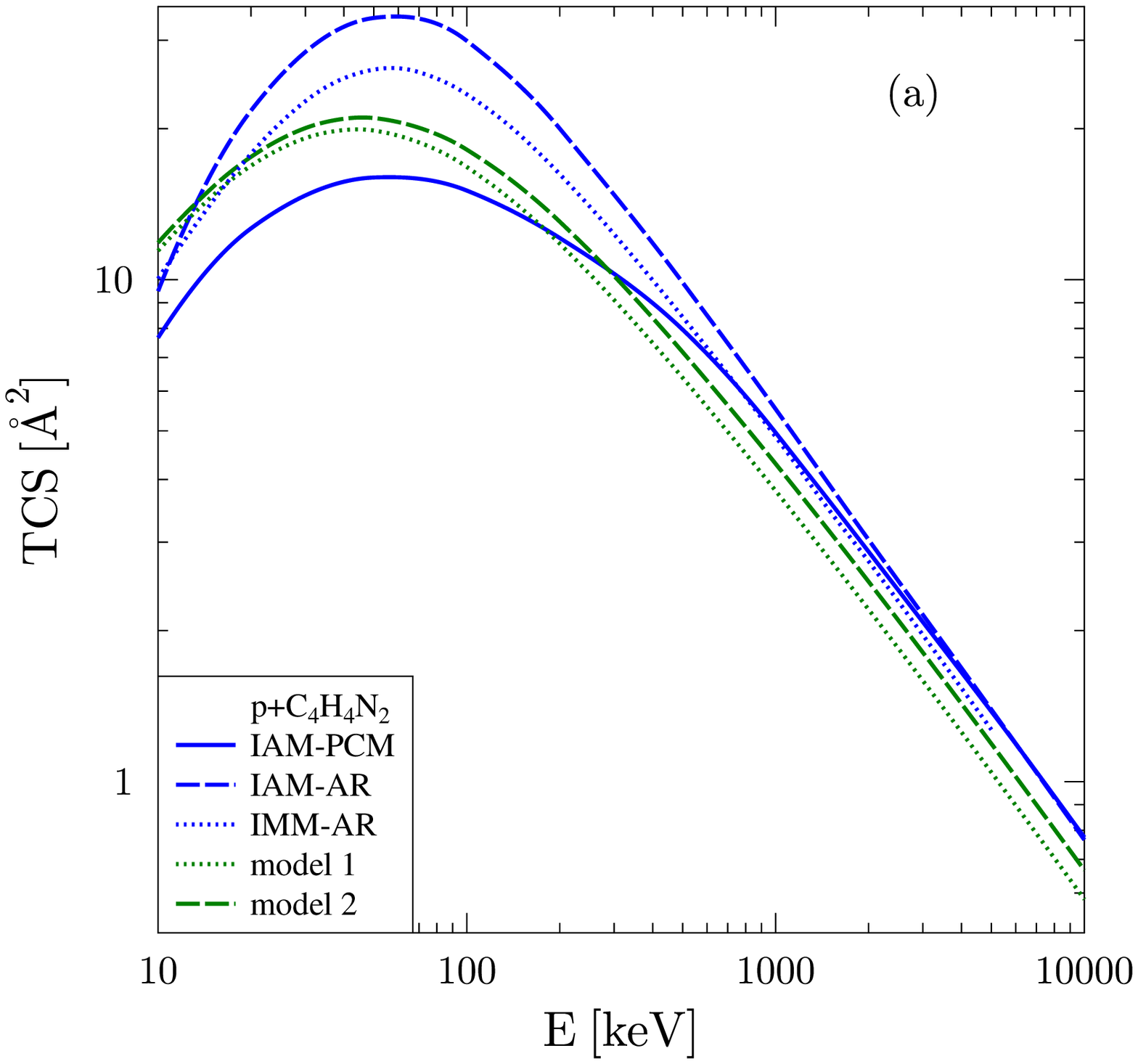}
}
\resizebox{0.49\textwidth}{!}{%
\includegraphics{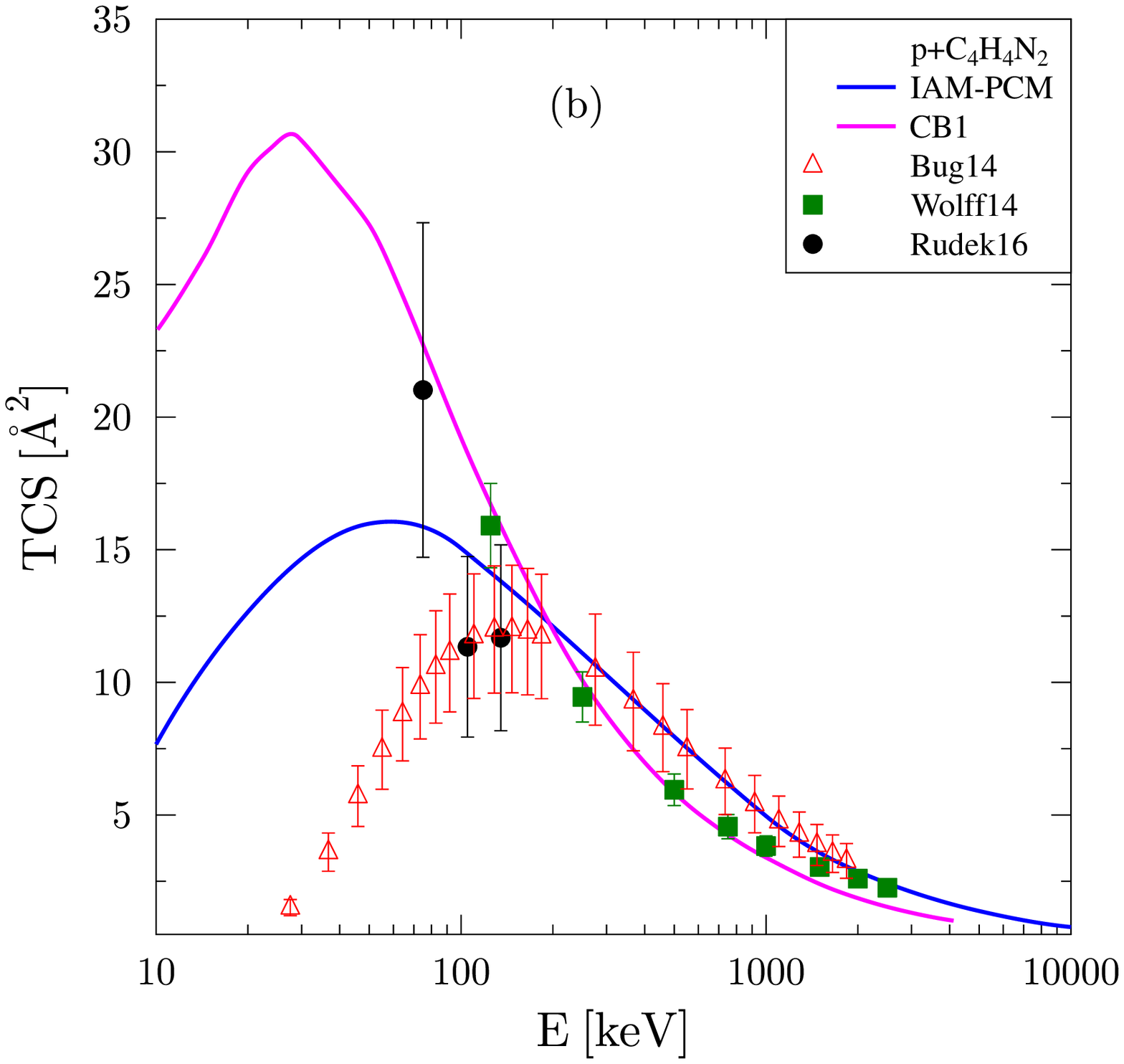}
}
\caption{%
Total net ionization cross section for proton-pyrimidine (C$_4$H$_4$N$_2$) collisions as a function
of impact energy on (a) a double-logarithmic scale and (b) a single-logarithmic scale.
IAM-PCM, IAM-AR: present calculations,
IMM-AR~\cite{Rudek16},
model~1 and model~2 denote model calculations based on equations~(\ref{eq:montescal1})
and (\ref{eq:montescal2}) as described in the text;
CB1: first Born approximation with corrected boundary conditions within
the CNDO approach~\cite{Rudek16};
experiments: Bug14~\cite{Bug14} (see also~\cite{Bug17}) for electron impact using
equivelocity conversion, 
Wolff14~\cite{Wolff14}, Rudek16~\cite{Rudek16} for proton impact.
}
\label{fig:py}
\end{center}
\end{figure*}

Figure~\ref{fig:py} shows the orientation-averaged net ionization cross section for
proton-pyrimidine collisions as a function of impact energy. In the left panel 
[figure~\ref{fig:py}(a)], 
we compare several model calculations on a double-logarithmic scale which emphasizes
the fall-off of the ionization cross section toward high impact energy.
This shows nicely how the present IAM-AR and IAM-PCM results merge 
in the $E\ge 1000$ keV range in which the atomic cross sections are so small that
no significant overlap occurs for {\it any} orientation. 
This is very different at lower energies: At the cross section
maximum around $E=60$ keV the IAM-AR cross section is about a factor of two
larger than its IAM-PCM counterpart. 

The two IAM calculations bracket the IMM-AR
results reported in~\cite{Rudek16} in most of the impact energy range shown. 
This is expected based on geometrical considerations.
On the one hand the (experimental) molecular cross sections
used in the IMM-AR to assemble the cross section for pyrimidine 
should be smaller than the zero-overlap IAM-AR predictions, but on the other
hand their sum should be larger than the IAM-PCM cross section for pyrimidine, since 
overlaps of the contributing molecular cross sections are neglected in the IMM-AR. 
The fact that the 
IMM-AR results become slightly smaller than the IAM-PCM calculations above
$E=1000$ keV cannot be explained on geometrical grounds. It points either
to an underestimation of the overlap effect by the IAM-PCM or to a small
inconsistency between the (atomic) TC-BGM results and the experimental cross
section data used in the IMM-AR calculation of~\cite{Rudek16}.
Given that no details about the latter were provided, we cannot 
offer a more definitive explanation for the (small) discrepancy.

The two model calculations 
included in figure~\ref{fig:py}(a)
are based on a semi-empirical scaling relation proposed by 
Montenegro and co-workers~\cite{Montenegro13}, and used
for proton-pyrimidine collisions in~\cite{Wolff14}.
According to that model the
net ionization cross section of an atom or molecule 
can be written as
\begin{equation}
  \sigma^{\rm net}(E) =
   \sum_k\sigma_k(E) = \sum_k \frac{Z_k\delta_k}{I_k^2}  F\left(\frac{E/M}{I_k} \right) 
\label{eq:montescal1}
\end{equation}
with the universal function
\begin{equation}
   F(x) =  \frac{A \ln (1+Bx)}{x} - \frac{AB}{(1+Cx)^4} .
\label{eq:montescal2}
\end{equation}
The sum in~(\ref{eq:montescal1}) is over the contributing atomic or molecular
orbitals, $Z_k$ is the occupation number
and $I_k$ the ionization potential (in atomic units) of the $k$th orbital, 
$\delta_k$ a parameter, 
and $E/M$ the collision
energy in keV/amu. The universal parameters in~(\ref{eq:montescal2}) are
$A=6.15 \times 10^3$, $B=7.0 \times 10^{-2}$, and $C=1.4 \times 10^{-2}$
to yield cross sections measured in units of $10^{-18}$ cm$^2$~\cite{Montenegro13}.

The model~1 calculation is identical to that reported in~\cite{Wolff14}. It uses
the
ionization potentials of the eleven most loosely bound, doubly-occupied 
($Z_k=2$) orbitals of pyrimidine quoted in~\cite{Wolff14} 
and $\delta_k=0.66$ for all $k$.
For model~2 we make the same choices for $Z_k$ and $\delta_k$, but use the
ionization potentials of the fifteen valence orbitals provided in~\cite{Rudek16}, which
are slightly different from those quoted in~\cite{Wolff14} for the outermost eleven.
Both model variants lead to similarly shaped cross section curves with the
model~2 results being larger than the model~1 results by 12--14 \%  
in the $E=300$ keV to $E=10$ MeV range in
which the curves appear as almost perfect straight lines on the double-logarithmic plot.
The difference in magnitude is mostly due to the fact that four additional orbitals are 
taken into account in model~2. 

We also carried out a model calculation based on the fifteen most loosely bound
\textit{atomic} orbitals using the
exchange-only DFT orbital energy eigenvalues on which the TC-BGM calculations are based. 
Given that the high-energy behavior of the model cross
section is controlled by
the first, Bethe-Born-like, term in equation~(\ref{eq:montescal2}) and the TC-BGM has
been shown to give results which agree very well with Bethe-Born predictions 
at high energies~(see section~\ref{sec:para} and~\cite{hjl18}),
one would expect that this model variant would agree with the IAM-PCM calculations at
least in the high-energy limit.
However, we found that 
the results of the atomic model are very similar to those of model~2 (which
is why they are not included in the figure) and, as a consequence, 
somewhat smaller than the IAM-PCM
cross section at $E\ge 300$ keV. 
We have checked that the high-energy discrepancy would essentially
be eliminated
if one would use $\delta_k=1.0$ instead of $\delta_k=0.66$ in the model. 
Such a choice would be consistent with the findings of~\cite{Montenegro13}:
In that paper $\delta=0.66$ was used to describe p-H collisions, while $\delta=1.0$ was
found to give excellent fits of the experimental ionization cross sections in the
p-N$_2$ and p-CH$_4$ systems. 
But even such an amended model would not agree with the IAM-PCM results at
lower impact energies, i.e., the latter are incompatible with
equations (\ref{eq:montescal1}) and (\ref{eq:montescal2}).
This suggests that in general one cannot expect very accurate results
when applying these equations
to collisions involving complex biomolecules.

Figure.~\ref{fig:py}(b) displays on a single-logarithmic plot the IAM-PCM
results together with experimental data for proton \cite{Wolff14, Rudek16} 
and equivelocity electron impact~\cite{Bug14}
and the results of a
first Born calculation with corrected boundary conditions (CB1) performed within the
CNDO approach~\cite{Rudek16}. The CB1 calculation gives smaller
cross section values than the IAM-PCM at energies above $E=200$ keV, in apparent agreement
with most of the experimental data points of~\cite{Wolff14}. However, the CB1
and experimental results differ in 
energy dependence above $E=1000$ keV. This is better seen on the double-logarithmic
plot provided in figure~5(a) of~\cite{Rudek16}. 
The IAM-PCM results agree very well with the electron-impact measurements of~\cite{Bug14} 
in the region above the cross section maximum in which
electron-impact data are expected to approach the proton-impact cross section.
Indeed, within combined error bars the electron-impact data
are in marginal agreement with the proton
measurements of~\cite{Wolff14}, although it appears as if the latter 
fall somewhat below the former. New experimental data with smaller error bars
would be needed to draw more definitive conclusions about the high-energy
behavior of the cross section.

The three experimental proton-impact data points of~\cite{Rudek16} at intermediate energies
have too large uncertainties to help shed light on the 
increasing deviations between the CB1 and IAM-PCM calculations in this region.
Given the first-order nature of the CB1 one would not expect this model to be
valid below $E=100$ keV, but in the absence of more experimental data the overall
situation remains unclear.

\begin{figure}
\begin{center}
\resizebox{0.49\textwidth}{!}{%
\includegraphics{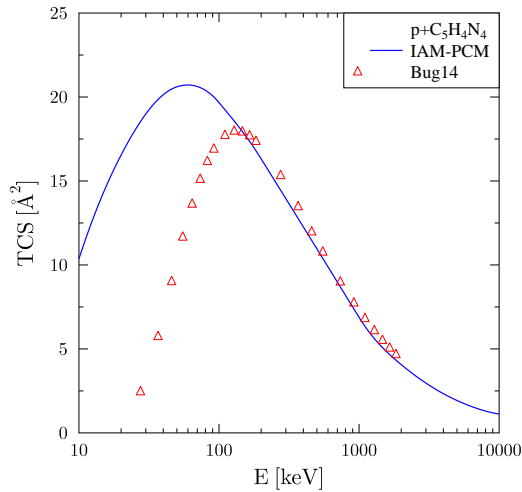}
}
\vspace{-1.5\baselineskip}
\caption{%
Total net ionization cross section for proton-purine (C$_5$H$_4$N$_4$) collisions as a function
of impact energy. 
IAM-PCM: present calculation;
experiments: Bug14~\cite{Bug14} (see also~\cite{Bug17}) for electron impact using
equivelocity conversion.
}
\label{fig:pu}
\end{center}
\end{figure}

Figure~\ref{fig:pu} shows IAM-PCM net ionization cross section results for proton-purine
collisions. We compare them with equivelocity electron-impact measurements from~\cite{Bug14} only, 
since we are not aware of other theoretical calculations 
or measurements for proton impact. Consistent with the pyrimidine case, the agreement
is excellent in the energy region above the experimental cross section maximum 
in which the sign of the projectile
charge is deemed to be unimportant.

\begin{figure}
\begin{center}
\resizebox{0.49\textwidth}{!}{%
\includegraphics{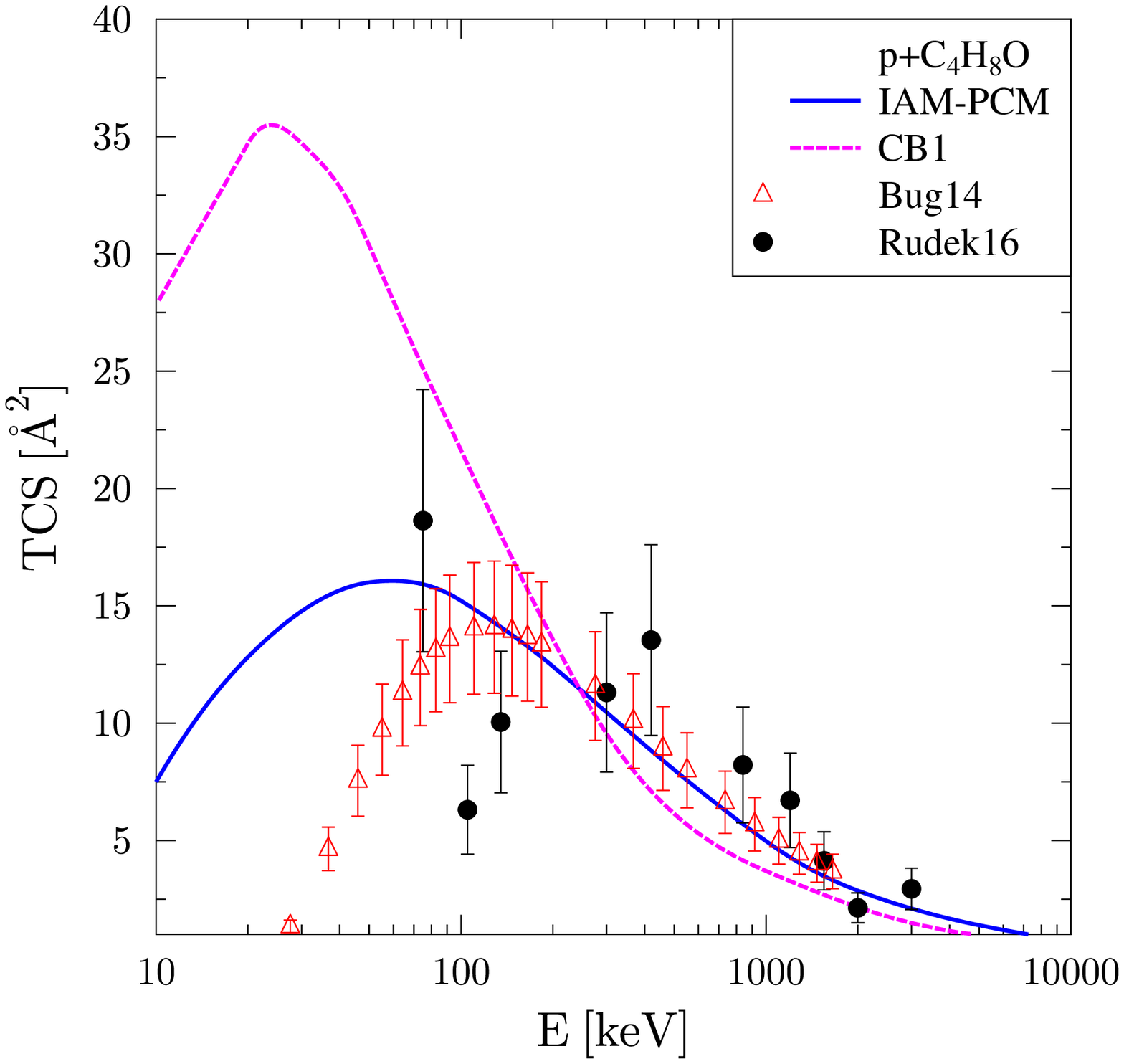}
}
\vspace{-1.5\baselineskip}
\caption{%
Total net ionization cross section for p-THF (C$_4$H$_8$O) collisions as a function
of impact energy. 
IAM-PCM: present calculation,
CB1: first Born approximation with corrected boundary conditions within
CNDO approach~\cite{Rudek16};
experiments: Bug14~\cite{Bug14} (see also~\cite{Bug17}) for electron impact using
equivelocity conversion, 
Rudek16~\cite{Rudek16} for proton impact.
}
\label{fig:thf}
\end{center}
\end{figure}

\begin{figure}
\begin{center}
\resizebox{0.49\textwidth}{!}{%
\includegraphics{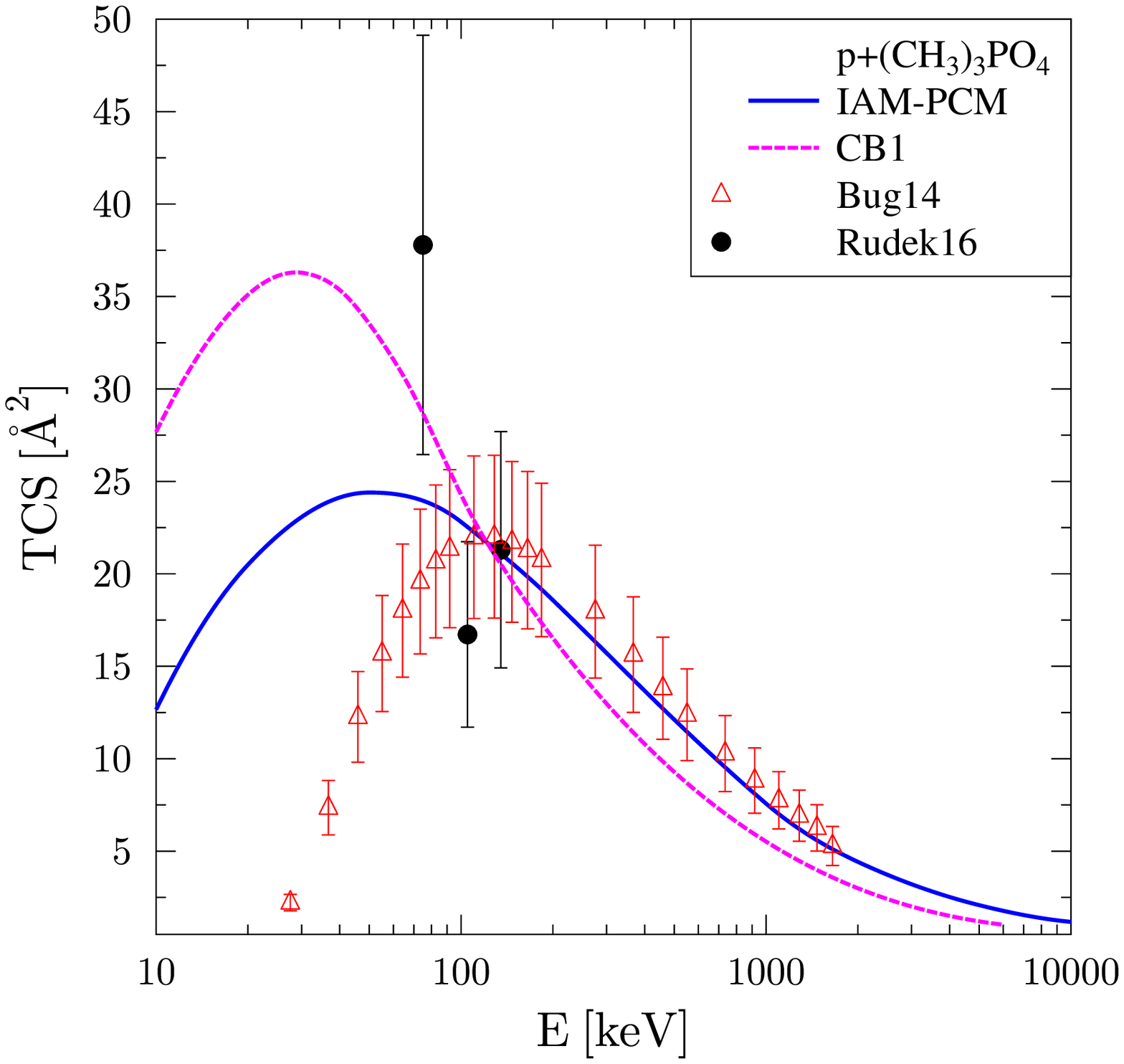}
}
\vspace{-1.5\baselineskip}
\caption{%
Total net ionization cross section for p-TMP ((CH$_3$)$_3$PO$_4$) collisions as a function
of impact energy. 
IAM-PCM: present calculation,
CB1: first Born approximation with corrected boundary conditions within
CNDO approach~\cite{Rudek16};
experiments: Bug14~\cite{Bug14} (see also~\cite{Bug17}) for electron impact using
equivelocity conversion, 
Rudek16~\cite{Rudek16} for proton impact.
}
\label{fig:tmp}
\end{center}
\end{figure}

Similar observations are made for the THF and TMP target molecules, as shown in
figures~\ref{fig:thf} and~\ref{fig:tmp}, respectively. For these two cases, in
addition to the electron data from~\cite{Bug14},  
proton-impact cross section measurements and
CB1 calculations from~\cite{Rudek16} are available for comparison.
As for the proton-pyrimidine case, the CB1 results
are somewhat lower than the IAM-PCM cross sections
at relatively high impact energies where first-order perturbation
theory is expected to be valid. They cross the IAM-PCM curve between $E=100$ and $E=200$ keV and
are probably too high at lower energies where the perturbation is too strong for a first-order
theory to be reliable. The measured data points of~\cite{Rudek16}
appear somewhat unsystematic and have error bars that are too large to differentiate
between the CB1 and IAM-PCM calculations. At high energies in particular, new measurements 
with smaller uncertainties are highly desirable to clarify the situation.
This caveat notwithstanding, the IAM-PCM results appear to 
agree with most experimental
data points for the four collision systems studied in this section. 
We thus feel encouraged to expand the application of
the model to a larger class of systems for which experimental data are sparse
and theoretical predictions are required.

\section{Scaling properties}
\label{sec:scaling}

Previous experimental and theoretical work provided some evidence that the (total) proton-impact
electron emission
cross sections of large biomolecules scale with the number of valence electrons~\cite{Iriki11a, Paredes15, Rudek16, Bug17}.
However, only a relatively small set of molecules has been investigated so far and it is not clear whether
the observed approximate scaling applies to a larger class of systems and can be used to predict
cross sections for experimentally inaccessible compounds. In this section we investigate this question 
by using the IAM-PCM to calculate proton-impact net ionization cross sections for four groups of
biologically relevant systems: pyrimidines, purines, amino acids, and nucleotides, 
the latter constituting the monomeric units of RNA and DNA~\cite{Adams81, Alberts15}. 
To our knowledge, for most
of the studied species the results
presented here are the first cross section data obtained from systematic, parameter-free calculations. 
Exceptions are the five DNA/RNA nucleobases 
cytosine, thymine, uracil, adenine and guanine. 
As mentioned in section~\ref{sec:validate}
the first three fall into the category of
pyrimidines, while the latter two are purines. For all of them 
classical~\cite{Lekadir09, Sarkadi15} and perturbative~\cite{Cappello08}
cross section calculations have been carried out, the latter in some cases within the CNDO
approach~\cite{Champion10, Galassi12, Champion13}.
A comparison with those earlier calculations, the IMM predictions of~\cite{Paredes15}, and the 
limited experimental data available~\cite{Tabet10b, Iriki11a, Iriki11b, Itoh13} will be presented elsewhere~\cite{hjl19}.

In the following, we consider three different scaling prescriptions. The first one uses the standard textbook definition
of valence electrons according to which all electrons in the outermost $n$-shells of the atoms that 
form the molecule under study
are included in the valence-electron count~\cite{Petrucci17}. In the second variant we only count the
{\it bonding} valence electrons, i.e., those electrons that form {\it lone pairs} are excluded. 
The scaled cross section is obtained by dividing the orientation-averaged IAM-PCM
result for a molecule with formula C$_{n_1}$H$_{n_2}$N$_{n_3}$O$_{n_4}$P$_{n_5}$ 
by the numbers 
\begin{equation}
N_{\rm VE} = 4n_1 + n_2 +5n_3+6n_4+5n_5
\label{eq:nve}
\end{equation}
and
\begin{equation}
N_{\rm BVE} = 4n_1 + n_2 +3n_3+2n_4+5n_5 ,
\label{eq:nbve}
\end{equation}
respectively, assuming in the latter case that all $L$-shell electrons of carbon participate in bonds (through hybridization), while
in nitrogen one and in oxygen two electron pairs do not.

\begin{figure}
\begin{center}
\resizebox{0.49\textwidth}{!}{%
\includegraphics{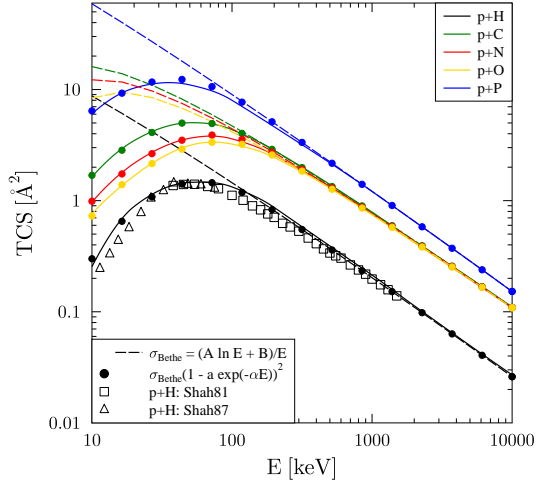}
}
\vspace{-1.5\baselineskip}
\caption{%
Total net ionization cross sections for proton collisions with atomic hydrogen (H), carbon (C), nitrogen (N),
oxygen (O), and phosphorus (P) as functions of impact energy. 
Full lines: present TC-BGM calculations,
dashed lines: Bethe-Born ionization cross sections $\sigma_{\rm Bethe} =
(A\ln E + B)/E$ with fit parameters $A$ and $B$ in appropriate units (see section~\ref{sec:para}),
$(\bullet )$: parametrizations discussed in section~\ref{sec:para}.
Experiments for p-H: Shah81~\cite{shah81}, Shah87~\cite{shah87}.
}
\label{fig:atomic}
\end{center}
\end{figure}

The third
prescription is based on the IAM and the observation that at high impact energies 
the net ionization cross sections
for p-C, p-N, and p-O collisions are very similar 
(i.e., $\sigma_X^{\rm net} \equiv \sigma_C^{\rm net} \approx \sigma_N^{\rm net} \approx \sigma_O^{\rm net}$) 
and to a good approximation four times larger
than the p-H cross section~\cite{hjl18}. This is demonstrated in figure~\ref{fig:atomic}, where the 
corresponding TC-BGM results are shown. Since the nucleotides studied further below contain
phosphorus, the p-P collision system is included in figure~\ref{fig:atomic} as
well. In this case, we find that the net ionization cross
section is approximately 1.5 times as large as that for carbon, nitrogen, and oxygen. 
Accordingly, the high-energy IAM prediction for the net ionization cross section of the molecule 
C$_{n_1}$H$_{n_2}$N$_{n_3}$O$_{n_4}$P$_{n_5}$ is to a good approximation 
$ (n_1+n_2/4+n_3+n_4+3 n_5/2) \sigma_X^{\rm net}$,
and we scale the (orientation-averaged) IAM-PCM cross sections in this variant by dividing them by 
the (fractional) coefficients
\begin{equation}
N_{\rm IAM} = n_1+\frac{n_2}{4}+n_3+n_4+ \frac{3n_5}{2}.
\label{eq:niam}
\end{equation}

\begin{figure*}
\begin{center}
\resizebox{0.49\textwidth}{!}{%
\includegraphics{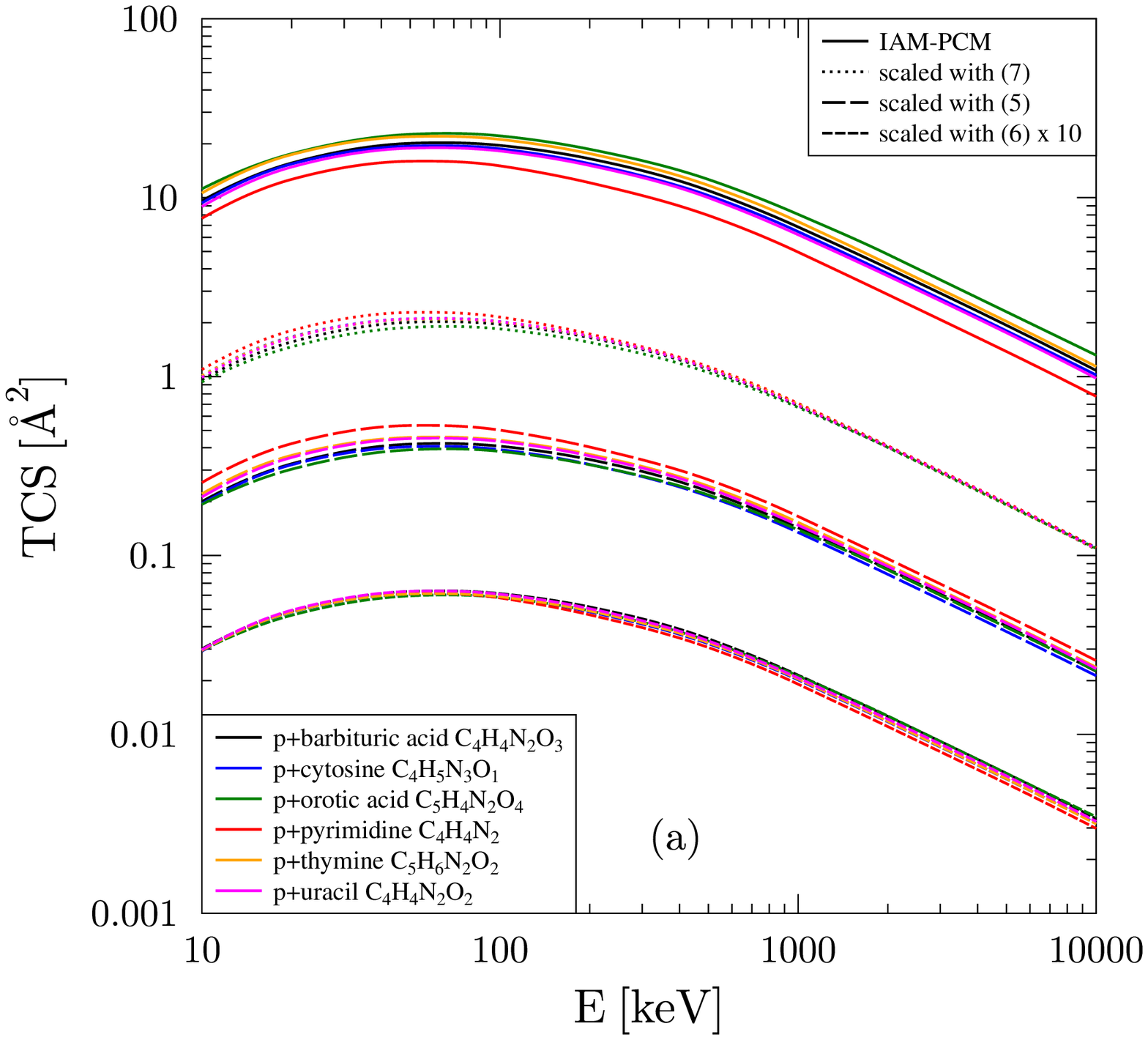}
}
\resizebox{0.49\textwidth}{!}{%
\includegraphics{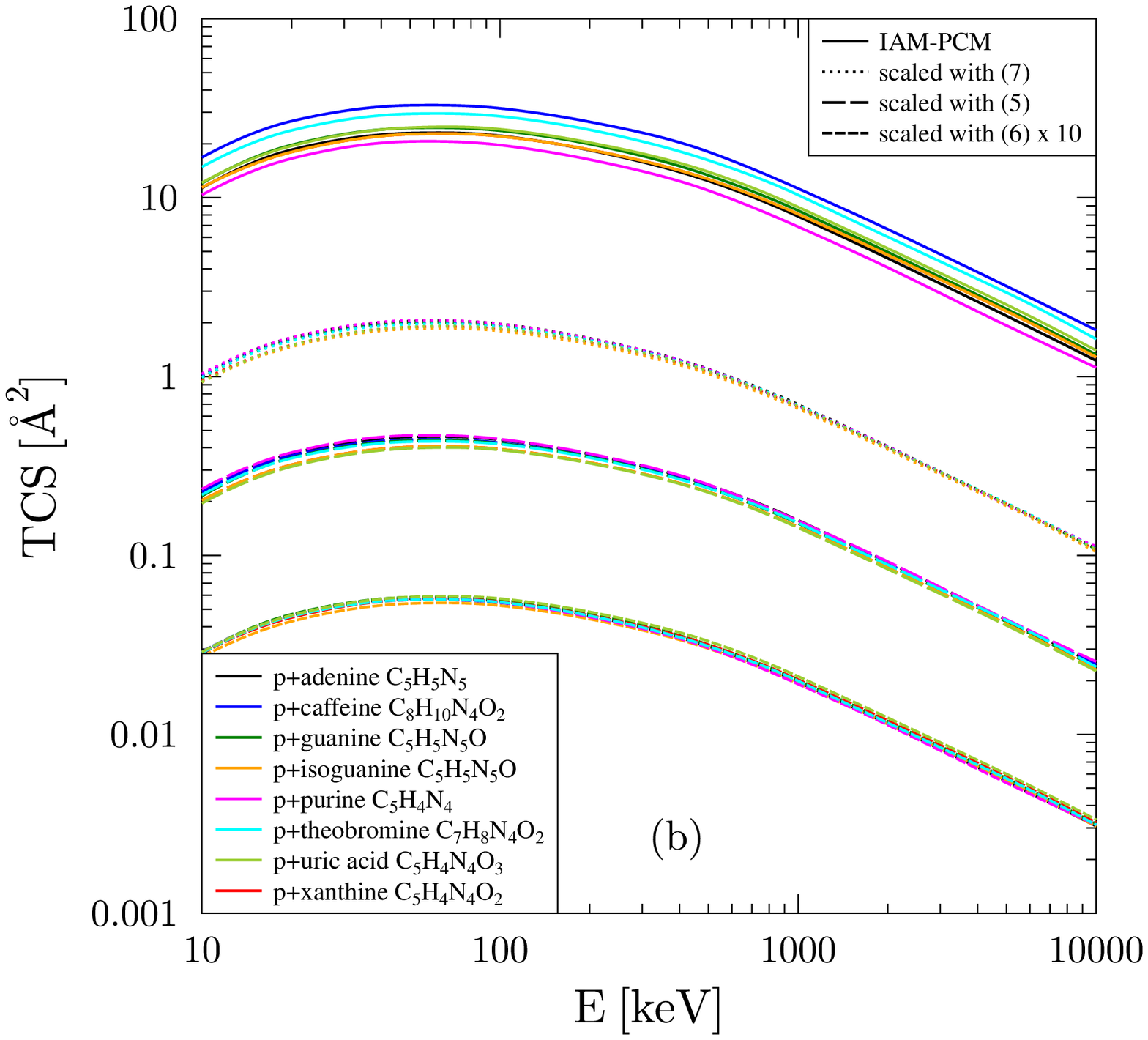}
}

\vspace{-1.5\baselineskip}

\resizebox{0.49\textwidth}{!}{%
\includegraphics{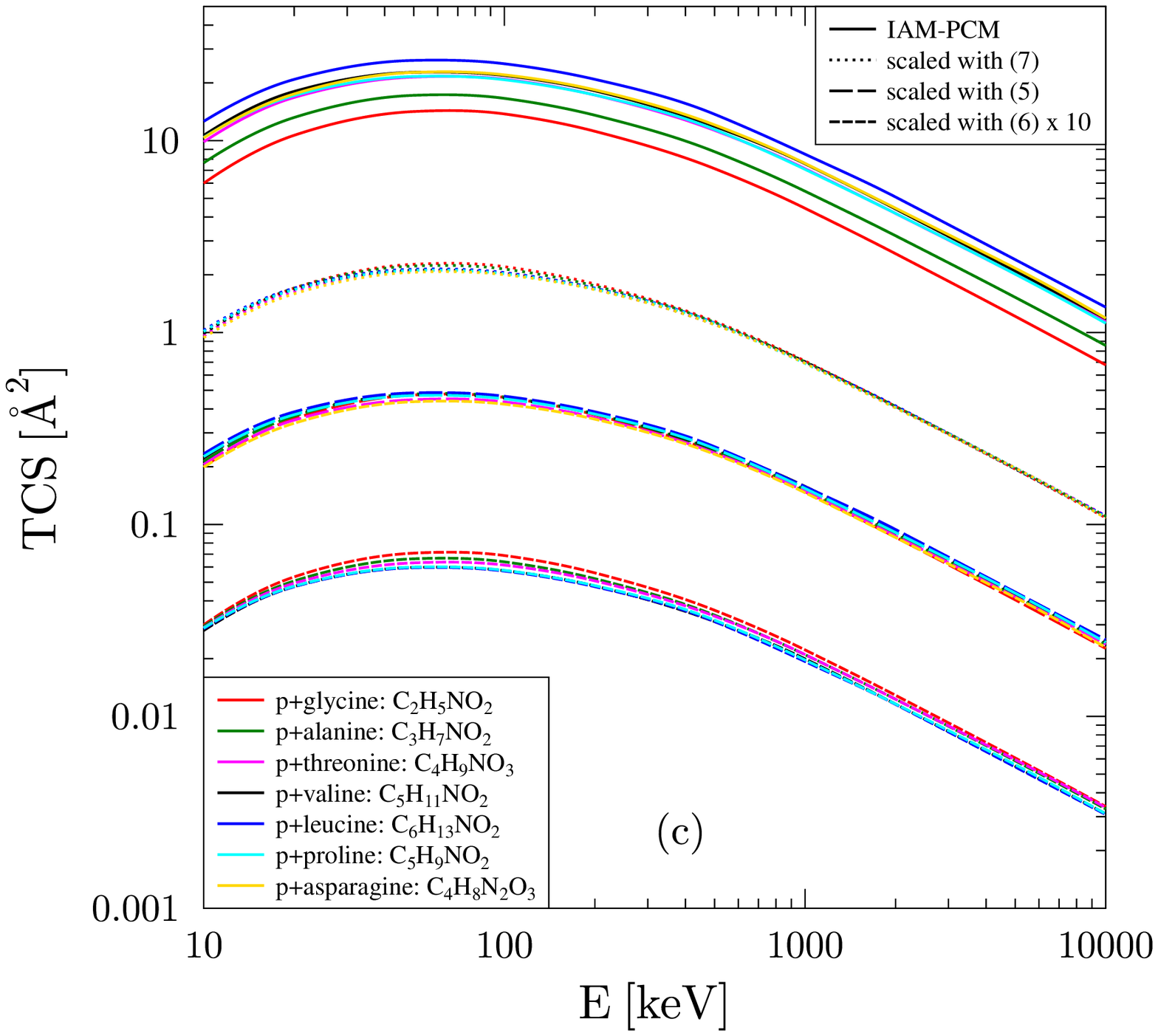}
}
\resizebox{0.49\textwidth}{!}{%
\includegraphics{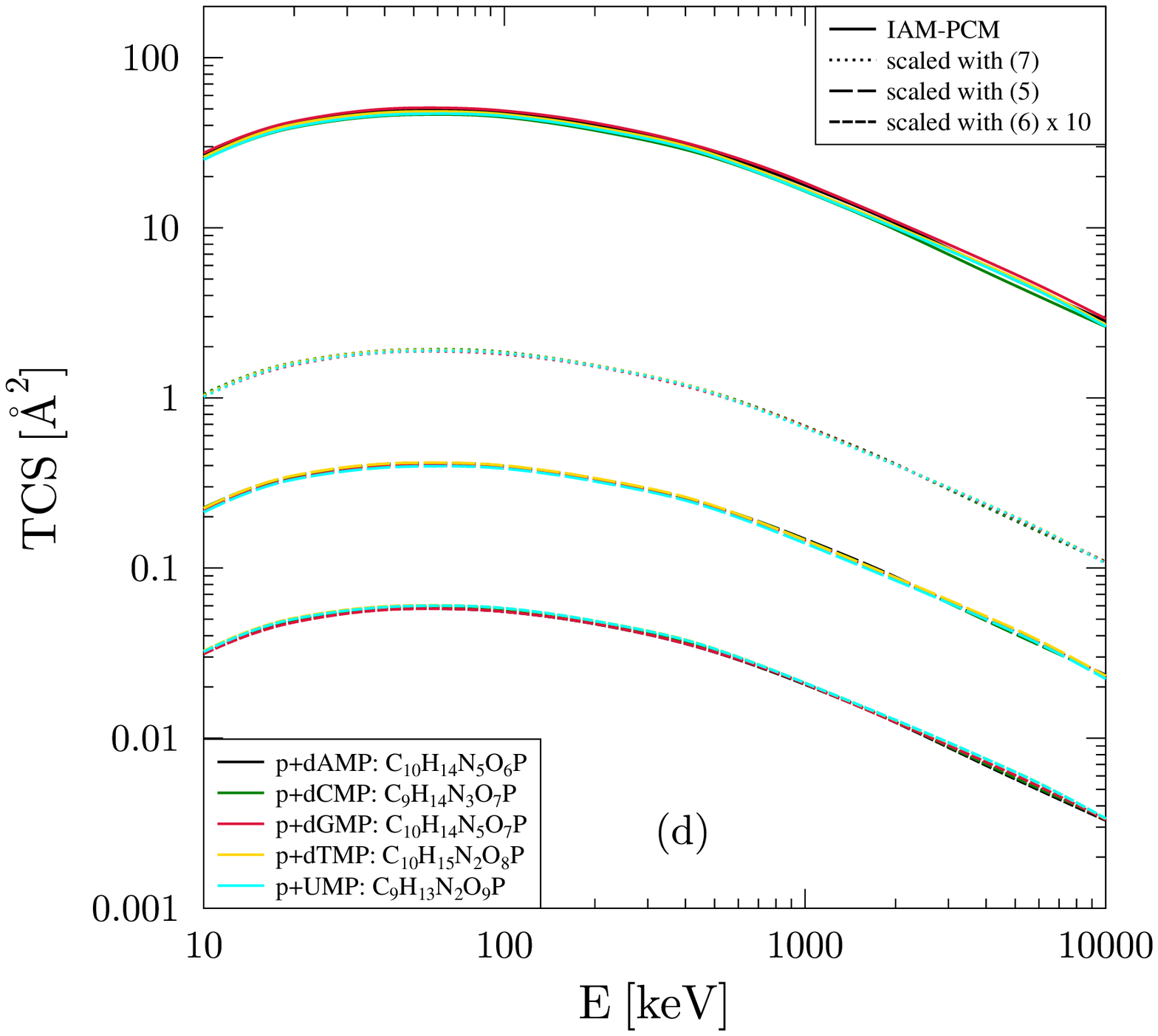}
}

\vspace{-1.5\baselineskip}
\caption{%
Total net ionization cross sections for proton collisions with 
(a) the pyrimidines barbituric acid, cytosine, orotic acid, pyrimidine, thymine, uracil;
(b) the purines adenine, caffeine, guanine, isoguanine, purine, theobromine, uric acid, xanthine;
(c) the amino acids glycine, alanine, threonine, valine, leucine, proline, asparagine;
(d) the DNA nucleotides 
deoxyadenosine monophosphate (dAMP), deoxycytidine monophosphate (dCMP),
deoxyguanosine monophosphate (dGMP), deoxythymidine monophosphate (dTMP),
and the RNA nucleotide uridine monophosphate (UMP)
as functions
of impact energy.
Full lines: present IAM-PCM calculations,
long-dashed lines: present IAM-PCM calculations divided by the number of valence electrons (\ref{eq:nve}),
short-dashed lines: present IAM-PCM calculations divided by the number of bonding valence 
electrons (\ref{eq:nbve}) and further divided by ten for visibility,
dotted lines: present IAM-PCM calculations divided by the IAM coefficients~(\ref{eq:niam}).
}
\label{fig:scal}
\end{center}
\end{figure*}

Figure~\ref{fig:scal} displays the scaled and unscaled cross section results for all systems studied in this section. The unscaled 
IAM-PCM cross sections are also provided in tabulated form in
the Appendix.
For the pyrimidines [figure~\ref{fig:scal}(a)] and purines [figure~\ref{fig:scal}(b)] the scaling with respect 
to the number of bonding valence electrons yields
better results than that with respect to the number of all valence electrons. For the amino acids [figure~\ref{fig:scal}(c)] the situation is reversed, most visibly so in the region around the cross section maximum, while for
the nucleotides [figure~\ref{fig:scal}(d)] both scaling procedures appear to work equally well. 

A more conclusive picture emerges when the cross sections are scaled by dividing them by the fractional
IAM coefficients (\ref{eq:niam}). In this case, we obtain for each group of systems a nearly universal curve.
Given the atomic results shown in figure~\ref{fig:atomic} this is to be expected at 
high energies where the IAM-PCM cross sections approach the IAM-AR limit. However, it is not obvious that the
scaling should also hold at lower energies where significant atomic cross section overlaps occur and the IAM-PCM calculations
are not orientation independent; e.g., we found that at $E=100$ keV the net ionization cross section of pyrimidine varies within a factor of two as
a function of orientation.
It is one of the main results of this work that despite these caveats the scaling holds if one accepts tolerances on the order
of 10\%.

\begin{figure}
\begin{center}
\resizebox{0.49\textwidth}{!}{%
\includegraphics{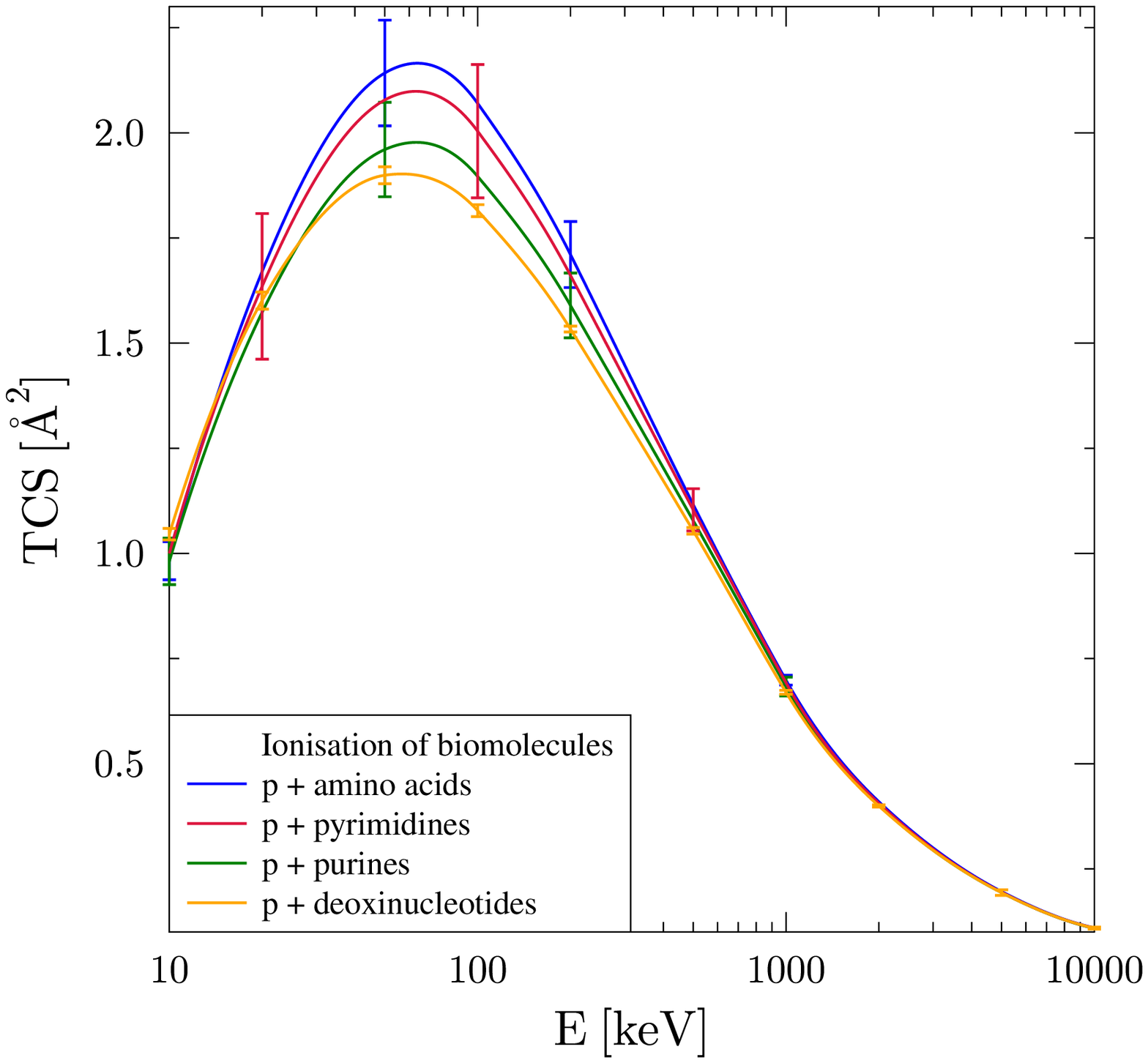}
}
\vspace{-1.5\baselineskip}
\caption{%
Average (IAM-based) scaled net ionization cross sections using (\ref{eq:niam}) for proton collisions with
pyrimidines, purines, amino acids, and DNA nucleotides
as functions of impact energy. 
The deviations of the actual results displayed in figure~\ref{fig:scal} from the averages are shown as error bars.
}
\label{fig:averages}
\end{center}
\end{figure}

Figure~\ref{fig:averages} provides a more differentiated view of the approximate universality of the IAM-based 
scaling. It shows on a linear
scale averages of the IAM-normalized cross sections and the deviations from these averages as error bars.
For the pyrimidines these deviations are as large as $\sim \pm 10$\%, while for the DNA nucleotides they are smaller
than $\pm 2$\%. The negligible spread of the latter can be traced back to the sugar-phosphate backbone, which is
the same for all DNA nucleotides, and, due to the large p-P cross section (cf. figure~\ref{fig:atomic}), 
is the main contributor to the total cross section.
Our results then indicate that the differences in the cross sections of the pyrimidines and purines are not relevant
for the ionization of a DNA molecule. Rather, they suggest that one can understand the latter as the ionization of one or another
of its nucleotide monomers, all of which fulfill the IAM-based scaling relation very accurately, i.e.,
it does not matter much which of them is actually ionized.

\section{Parametrization}
\label{sec:para}

\begin{figure}
\begin{center}
\resizebox{0.49\textwidth}{!}{%
\includegraphics{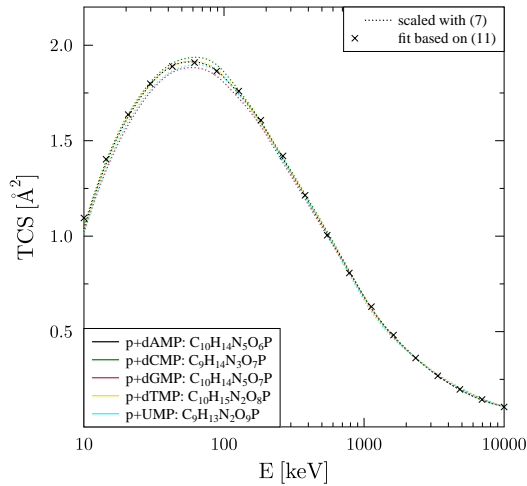}
}
\vspace{-1.5\baselineskip}
\caption{%
IAM-based scaled net ionization cross sections using (\ref{eq:niam}) for proton collisions with
the DNA nucleotides 
deoxyadenosine monophosphate (dAMP), deoxycytidine monophosphate (dCMP),
deoxyguanosine monophosphate (dGMP), deoxythymidine monophosphate (dTMP),
and the RNA nucleotide uridine monophosphate (UMP)
as functions of impact energy. 
Crosses: fit based on~(\ref{eq:para2}) using the parameters $A=135.0$ \AA$\!^2$keV , $B=-190.0$ \AA$\!^2$keV, 
$a=0.965$ and $\alpha = 0.102$ keV$^{-1/2}$.
}
\label{fig:para}
\end{center}
\end{figure}

The finding that the IAM-based scaling works very well for a large class of systems raises the question
whether the proton-impact electron emission cross sections of biomolecules can be parametrized in a 
convenient way. 
To address this question we re-inspect figure~\ref{fig:atomic} which in addition to the atomic net ionization
cross sections obtained from the TC-BGM calculations displays Bethe-Born results in which the
parameters $A$ and $B$ in the cross section formula
\begin{equation}
 \sigma_{\rm Bethe}(E) = \frac{A\ln E}{E} + \frac{B}{E} 
\end{equation}
were determined by a fitting procedure using the Fano representation~\cite{hjl18}. The agreement is excellent for impact energies
above $E\approx 200$ keV. At lower energies, the TC-BGM cross sections are smaller than the Bethe-Born
predictions, mostly because electron capture (which is described by the TC-BGM) gains importance and
ultimately takes over as the dominant electron removal channel. One may argue that the occurrence of
capture effectively decreases the projectile charge $Q_P$, thereby reducing the amount of direct
ionization to the continuum, which in the Bethe-Born approximation is proportional to the square of $Q_P$~\cite{inokuti71}.
This suggests the ansatz
\begin{equation}
  \sigma^{\rm ion}(E) = [Q_{P,\rm eff}(E)]^2 \sigma_{\rm Bethe}(E)
\label{eq:para1}
\end{equation}
with an effective charge of the form
\begin{equation}
  Q_{P,\rm eff}(E) = 1 - a \exp(-\alpha E) .
\label{eq:qpeff}
\end{equation}
We note that similar parametrizations have been proposed in the past to model the electron loss from neutral 
projectiles (see~\cite{Gervasoni96} and references therein).
The full circles in figure~\ref{fig:atomic} show the results obtained from equations~(\ref{eq:para1}) 
and (\ref{eq:qpeff}) with the parameters 
listed in Table~\ref{tab:a_alpha}. The agreement with the TC-BGM cross section curves is almost perfect.
A similar parametrization should then work for biomolecular targets. However, in this case we
found that the overlap effects taken into account in the IAM-PCM result in a flatter shape of the cross section
curves compared to their atomic counterparts and the modified formula
\begin{equation}
  \sigma^{\rm ion}_{\rm mol}(E) = [1-a\exp(-\alpha \sqrt{E})]^2 \sigma_{\rm Bethe}(E)
\label{eq:para2}
\end{equation}
provides better fits. This is demonstrated for the nucleotides in figure~\ref{fig:para}.
Formula~(\ref{eq:para2}) agrees with the orientation-averaged IAM-PCM calculations in the entire
impact energy range from $E=10$ keV to
$E=10$ MeV to within $\pm 3$\%.

\begin{table}
\caption{
\label{tab:a_alpha}
 Parameters $\alpha$ (in keV$^{-1}$) and $a$ used in~(\ref{eq:qpeff}) to model the effective charges 
for the atomic targets hydrogen (H), 
carbon (C), nitrogen (N), oxygen (O), and phosphorus (P). 
}
\begin{tabular}{cccccc}
\hline \hline
 & H & C & N & O & P \\
\hline
 $\alpha$ & 0.030 & 0.033 & 0.024 & 0.021 & 0.040 \\
 a & 1.10 & 0.94 & 0.91 & 0.87 & 1.00 \\
\hline \hline
\end{tabular}
\end{table}

\section{Concluding remarks}
\label{sec:conclusions}

We have used the IAM-PCM, introduced in recent work, to calculate proton-impact net ionization
cross sections for a large class of biologically relevant molecules from $E=10$ keV to $E=10$ MeV impact energy.
We have found overall good agreement with the limited experimental data available for pyrimidine, purine, THF,
and TMP and have made predictions for a number of larger systems
including amino acids and nucleotides. To our knowledge, the results reported for the latter are the first
cross sections obtained from a parameter-free theoretical model. 
It is shown that they follow a scaling rule which is based on IAM-derived fractional coefficients
and can be represented by a simple analytical formula to within 3\% accuracy. 
Scaling prescriptions based on the
number of (bonding) valence electrons which were advocated in previous works yield less conclusive results.

The IAM-PCM is based on a geometrical interpretation of a molecular total cross section as the combined area of
overlapping circles which represent atomic cross sections. 
The latter are calculated from first principles
using accurate atomic potentials and the nonperturbative TC-BGM for orbital
propagation. Once these cross sections have been computed, the IAM-PCM procedure to assemble the
molecular cross section is computationally inexpensive and numerically accurate. 

We envision that the method can be used to describe collisions involving long-chain molecules and polymers,
such as peptides and large DNA sections, 
in terms of cross section calculations for (clusters of) amino acids and nucleotides.
Further work in this direction is in progress. 
Future studies will also be concerned with the electron capture channel and with the extraction of
charge-state-correlated multiple ionization data. The latter will be particularly relevant
for collision systems involving multiply-charged projectile ions.

\begin{acknowledgments}
This work was supported by the Natural Sciences and Engineering Research Council of Canada (NSERC).
One of us (H. J. L.) would like to thank the Center for Scientific Computing, University of Frankfurt 
for making their
High Performance Computing facilities available.
\end{acknowledgments}

%
\bibliography{pcm-epjd2018}

\appendix

\section{Net ionization cross sections}
\label{sec:appendixv}
In this appendix we present tables with the orientation-averaged IAM-PCM net ionization cross sections for all systems 
studied in section~\ref{sec:scaling}.

\begin{table}[h]
 \caption{Orientation-averaged IAM-PCM net ionization cross sections for proton collisions with pyrimidines (in \AA$^2$).}
 \label{tab:Pyr2}
 \begin{tabular}{rcccccc}
  \hline \hline
  E [keV] &Barbituric acid &  Cytosine   & Orotic acid & Pyrimidine  &  Thymine    & Uracil     \\
          &  C$_4$H$_4$N$_2$O$_3$ &  C$_4$H$_5$N$_3$O  & C$_5$H$_4$N$_2$O$_4$  &  C$_4$H$_4$N$_2$ & C$_5$H$_6$N$_2$O$_2$ 
          &  C$_4$H$_4$N$_2$O$_2$ \\ 
  \hline
   10 	  & 9.62	& 9.34	& 11.19	& 7.67	& 10.62	& 8.92 \\	
   20 	  & 15.62	& 15.42	& 17.60	& 12.66	& 17.41	& 14.83 \\	
   50 	  & 20.12	& 19.44	& 22.60	& 15.98	& 21.93	& 18.90 \\	
  100 	  & 19.59	& 18.75	& 22.14	& 15.05	& 21.17	& 18.23 \\	
  200 	  & 16.50	& 15.49	& 18.64	& 12.11	& 17.63	& 15.13 \\	
  500 	  & 10.95	& 10.28	& 12.64	& 7.96	& 11.72	& 9.99 \\	
 1000 	  & 6.86	& 6.46	& 8.07	& 4.96	& 7.35	& 6.21 \\	
 2000 	  & 4.03	& 3.76	& 4.82	& 2.87	& 4.26	& 3.65 \\	
 5000 	  & 1.94	& 1.80	& 2.30	& 1.38	& 2.03	& 1.76 \\	
10000 	  & 1.08	& 1.02	& 1.32	& 0.77	& 1.14	& 0.98 \\              
  \hline \hline
 \end{tabular}
\end{table}
\begin{table}[h]
 \caption{Orientation-averaged IAM-PCM net ionization cross sections for proton collisions with purines (in \AA$^2$).}
 \label{tab:Purin2}
 \begin{tabular}{rccccccc}
  \hline \hline
  E [keV] &  Adenine    & Caffeine    &   Guanine   &   Purine    & Theobromine & Uric Acid   &  Xanthine   \\
          &  C$_5$H$_5$N$_5$   & C$_8$H$_{10}$N$_4$O$_2$ & C$_5$H$_5$N$_5$O  & C$_5$H$_4$N$_4$ 
          &  C$_7$H$_8$N$_4$O$_2$   &  C$_5$H$_4$N$_4$O$_3$  &  C$_5$H$_4$N$_4$O$_2$  \\
  \hline
   10 	  & 11.32 & 16.80 & 12.04 & 10.36 & 14.86 & 12.13 & 11.40 \\	
   20 	  & 18.53 & 26.78 & 19.70 & 16.55 & 23.77 & 19.47 & 18.00 \\	
   50 	  & 22.96 & 32.82 & 24.52 & 20.61 & 29.42 & 24.63 & 22.64 \\	
  100 	  & 22.17 & 31.56 & 23.68 & 19.67 & 28.49 & 24.11 & 22.03 \\	
  200 	  & 18.41 & 26.52 & 19.91 & 16.32 & 23.94 & 20.32 & 18.53 \\	
  500 	  & 12.40 & 18.04 & 13.33 & 10.96 & 16.18 & 13.94 & 12.68 \\	
 1000 	  & 7.86  & 11.25 & 8.47  & 6.88  & 10.35 & 8.86  & 8.10  \\	
 2000 	  & 4.58  & 6.66  & 4.96  & 4.07  & 6.03  & 5.20  & 4.78  \\	
 5000 	  & 2.18  & 3.20  & 2.39  & 1.93  & 2.95  & 2.52  & 2.32  \\	
10000 	  & 1.23  & 1.82  & 1.33  & 1.12  & 1.62  & 1.40  & 1.28  \\              
  \hline \hline
 \end{tabular}
\end{table}
\begin{table}[h]
 \caption{Orientation-averaged IAM-PCM net ionization cross sections for proton collisions with amino acids (in \AA$^2$).}
 \label{tab:Amino2}
 \begin{tabular}{rccccccc}
  \hline \hline
  E [keV] &  Alanine    & Asparagine  &   Glycine   &   Leucine   &   Proline   & Threonine   &  Valine     \\
          & C$_3$H$_7$NO$_2$   & C$_4$H$_8$N$_2$O$_3$ & C$_2$H$_5$NO$_2$   & C$_6$H$_{13}$NO$_2$& C$_5$H$_9$NO$_2$   
          & C$_4$H$_9$NO$_3$   & C$_5$H$_{11}$NO$_2$ \\
  \hline
   10 	  & 7.66  & 10.34 & 5.99  & 12.59 & 10.39 & 9.85  & 10.64 \\	
   20 	  & 13.20 & 17.51 & 10.66 & 20.85 & 17.17 & 16.78 & 18.04 \\	
   50 	  & 17.22 & 22.67 & 14.17 & 26.18 & 21.58 & 21.44 & 22.69 \\	
  100 	  & 16.60 & 22.01 & 13.77 & 25.14 & 20.75 & 20.88 & 21.83 \\	
  200 	  & 13.56 & 18.38 & 11.18 & 20.86 & 17.28 & 17.28 & 18.17 \\	
  500 	  & 8.76  & 12.08 & 7.14  & 13.72 & 11.43 & 11.32 & 11.86 \\	
 1000 	  & 5.45  & 7.61  & 4.44  & 8.50  & 7.08  & 7.14  & 7.57  \\	
 2000 	  & 3.19  & 4.45  & 2.57  & 5.08  & 4.18  & 4.17  & 4.40  \\	
 5000 	  & 1.53  & 2.16  & 1.21  & 2.40  & 2.01  & 2.02  & 2.09  \\	
10000 	  & 0.86  & 1.19  & 0.68  & 1.35  & 1.12  & 1.14  & 1.18  \\              
  \hline
 \end{tabular}
\end{table}

\begin{table}[h]
 \caption{Orientation-averaged IAM-PCM net ionization cross sections for proton collisions with the nucleotides
          deoxyadenosine monophosphate (dAMP), deoxycytidine monophosphate (dCMP),
          deoxyguanosine monophosphate (dGMP), deoxythymidine monophosphate (dTMP),
          uridine monophosphate (UMP) (in \AA$^2$).}
 \label{tab:Nukl2}
 \begin{tabular}{rccccc}
  \hline \hline
  E [keV]      &   dAMP      &   dCMP      &   dGMP      &    dTMP     &    UMP      \\
               & C$_{10}$H$_{14}$N$_5$O$_{6}$  & C$_{9}$H$_{14}$N$_3$O$_{7}$P  & C$_{10}$H$_{14}$N$_5$O$_{7}$P 
               & C$_{10}$H$_{15}$N$_2$O$_{8}$P & C$_{9}$H$_{13}$N$_2$O$_{9}$P  \\
  \hline
   10 	& 27.01 & 25.27  & 27.53  & 26.03 & 25.11 \\	
   20 	& 41.75 & 38.86  & 42.22  & 40.61 & 39.15 \\	
   50 	& 49.63 & 46.22  & 50.68  & 48.22 & 46.69 \\	
  100 	& 47.63 & 44.63  & 48.72  & 46.43 & 45.30 \\	
  200 	& 40.34 & 37.35  & 41.34  & 39.16 & 38.13 \\	
  500 	& 27.55 & 25.73  & 28.27  & 26.89 & 26.13 \\	
 1000 	& 17.76 & 16.40  & 18.33  & 16.90 & 16.48 \\	
 2000 	& 10.64 & 9.75   & 10.92  & 10.20 & 9.97  \\	
 5000 	& 4.94  & 4.57   & 5.33   & 5.07  & 4.94  \\	
10000 	& 2.82  & 2.62   & 2.92   & 2.69  & 2.63  \\              
  \hline \hline
 \end{tabular}
\end{table}

\end{document}